# On the application of Canonical Perturbation Theory up to the dissociation threshold

Sahin BUYUKDAGLI and Marc JOYEUX[(a)]

*Laboratoire de Spectrométrie Physique (CNRS UMR 5588),*

*Université Joseph Fourier Grenoble 1,*

*BP 87, 38402 St Martin d'Hères Cedex, France*

**Abstract :** We investigate a model system consisting of a Morse oscillator strongly coupled to a doubly-degenerate bending degree of freedom and show that Canonical Perturbation Theory is able to provide a fairly precise, though not exact, approximation of the Hamiltonian up to the dissociation threshold. Quantum mechanical results and classical ones are discussed in this Letter.

(a) email : Marc.JOYEUX@ujf-grenoble.fr

We have recently shown how canonical perturbation theory (CPT) can be adapted to investigate the highly excited dynamics of semi-rigid molecules, floppy molecules (i.e. molecules with several equilibrium positions), and non-adiabatic systems (i.e. molecules with several crossing diabatic electronic surfaces) : For a recent review, see for example Ref. [1] and references therein. Whether based on classical mechanics, like the Birkhoff-Gustavson method [2] and the Lie transform one [3], or quantum mechanics, like the Van-Vleck procedure [4], the purpose of CPT is always to rewrite the Hamiltonian of the investigated system in terms of as complete as possible a set of classical constants of the motion or good quantum numbers. The transformed Hamiltonian, whose dynamics can usually be analyzed in great detail thanks to the conserved quantities, then provides the keys for understanding the principal properties of the initial system. For example, we have used CPT to characterize the bifurcations of semi-rigid molecules and the isomerization dynamics of floppy molecules [5,6], to detect monodromy, a rather newly described mathematical concept [7], in floppy molecules [8,9] as well as in $CO_2$ [10], to understand the effect of the $X^2A_1$-$A^2B_2$ conical intersection on the vibronic spectrum of $NO_2$ and imagine an efficient method to adjust the coefficients of an effective Hamiltonian which describes this intersection [11], and to define meaningful local modes for the study of intramolecular vibrational energy redistribution in floppy molecules like $H_2O_2$ [12]. The purpose of the present Letter is to answer the following question : Up to what energy is CPT expected to describe correctly a molecule with a dissociation threshold ? Indeed, it is well known, that a Morse oscillator $H=\frac{1}{2M}p_R^2+D(1-e^{-\alpha(R-R_0)})^2$ is formally equivalent, up to the dissociation threshold, to an anharmonic oscillator $H=\omega I+xI^2$, where $I$ is the action integral of the harmonic oscillator ($I=\hbar(n+\frac{1}{2})$ in the semiclassical limit), $\omega=(2D\alpha^2/M)^{1/2}$ its fundamental frequency, and $x=-\omega^2/(4D)$ the first anharmonicity. This result is obtained when solving analytically Schrödinger's equation [13], but also when performing second order CPT [14] (higher orders lead to null corrections). A Morse oscillator is therefore perfectly well described by CPT up to the threshold. But what about a Morse oscillator coupled to another degree of freedom ? To our knowledge, this point has not been studied in detail and it is the goal of this Letter to provide an indication of what can be expected with this respect. More precisely, we investigated a model system consisting of a Morse oscillator strongly coupled to a doubly-degenerate bending degree of freedom and showed that CPT is still able to provide a fairly precise, though not exact, approximation of the Hamiltonian up to the dissociation threshold. Quantum mechanical results and classical ones will successively be briefly discussed in the remainder of this article.

The Hamiltonian we studied describes the vibrations of a linear triatomic molecule expressed in Jacobi coordinates, with the non-dissociating stretching degree of freedom frozen at equilibrium. It writes

$$H=\frac{1}{2M}p_R^2+\frac{1}{2}\left(\frac{1}{MR^2}+\frac{1}{mr_0^2}\right)\left(p_\gamma^2+\frac{p_\varphi^2}{\sin^2\gamma}\right)+D\left(1-e^{-\alpha(R-R_0)}\right)^2+C(1-\cos\gamma)e^{-\alpha(R-R_0)} \tag{1}$$

where $R$ is the Morse oscillator coordinate and $p_R$ its conjugate momentum, $\gamma$ the Jacobi bending angle and $p_\gamma$ its conjugate momentum, $p_\varphi$ the vibrational angular momentum arising from the degeneracy of the bending motion, and $MR^2$ and $mr_0^2$ the moments of inertia associated with the Morse oscillator and the frozen stretch degree of freedom, respectively. The last term in the right-hand side of Eq. (1) describes a realistic stretch-bend interaction potential, in the sense that the bend behaves like an harmonic oscillator close to the bottom $(R,\gamma)=(R_0,0)$ of the potential energy surface (PES), while the potential no longer depends on $\gamma$ at very large $R$ stretches (cf. Fig. 1). Moreover, the geometry of the PES depends crucially on the relative values of $C$ and $D$ : Indeed, if $C \le D$, there exists a saddle at $\gamma = \pi$, which is located below the dissociation threshold, while there is no such saddle in the case where $C > D$ (cf. Fig. 1). Numerical values used throughout this work are α=2.5 $Å^{-1}$, $R_0$ =2.5 Å, $mr_0^2$ =1.6948 $10^{-27}$ kg.$Å^2$, $MR_0^2$ =8.0370 $10^{-27}$ kg.$Å^2$, $D$=15000 $cm^{-1}$, and $C$=5625 $cm^{-1}$ or $C$=18750 $cm^{-1}$. These values result in a stretch fundamental frequency of about 2700 $cm^{-1}$ and a bend frequency of about 450 $cm^{-1}$ ($C$=5625 $cm^{-1}$) or 830 $cm^{-1}$ ($C$=18750 $cm^{-1}$).

Exact eigenvalues and eigenvectors of the Hamiltonian of Eq. (1) were first obtained by expanding $H$ up to very high order (300) with respect to the $R$ coordinate in the neighbourhood of $R$= $R_0$ and diagonalizing the Hamiltonian matrix of size 3000*3000 built in the direct product basis of the rotating Morse oscillator and the free rotor. 98 rotationless ($J = \ell = 0$) bound states with up to 40 quanta of excitation in the bend degree of freedom and 8 quanta in the stretch were obtained for $C$=5625 $cm^{-1}$, and 48 rotationless bound states with up to 20 quanta of excitation in the bend and 7 quanta in the stretch for $C$=18750 $cm^{-1}$. All states are converged to better than 0.01 $cm^{-1}$. The list of assigned states can be obtained by sending an email to the authors.

CPT calculations were then performed, in order to compare exact and perturbative results. The procedure to use, as well as the form of the transformed Hamiltonian, actually depend on the relative values of $C$ and $D$. Indeed, if $C \le D$, wave functions may be delocalized over the whole range $0 \le \gamma \le \pi$, because of the saddle at $\gamma = \pi$. In this case, the procedure we derived for floppy molecules [1,6] must be used. The first step of this procedure consists in expanding the Hamiltonian in the neighbourhood of a Minimum Energy Path (MEP) in Taylor series with respect to the stretch coordinate and in Fourier series with respect to the bend coordinate. After transformation of the stretch coordinates into dimensionless creation and annihilation operators $(a_1^+, a_1)$, the expanded Hamiltonian writes

$$H = \sum_{k,m,M,P,N} h_{kmMPN} (a_1^+)^k \, a_1^m (\cos\gamma)^M \sigma^P (J^2)^N \qquad (2)$$

where the $h_{kmMPN}$ are real coefficients, and $P$ is equal to 0 or 1. The operator $\sigma = \sin\gamma \frac{\partial}{\partial\gamma}$ arises from the non-commutativity of $\cos\gamma$ and $J^2 = -\frac{1}{\sin\gamma}\frac{\partial}{\partial\gamma}\sin\gamma\frac{\partial}{\partial\gamma} - \frac{1}{\sin^2\gamma}\frac{\partial^2}{\partial\varphi^2}$. A series of canonical transformations $H \to \exp(S) H \exp(-S)$ is then applied, where $S$ is chosen such as to cancel the terms with $k \neq m$ up to increasingly higher values of $k+m$. After a certain number of transformations, called the "order" of the perturbation series, the remaining terms with $k \neq m$, which have hopefully become very small in the investigated region of the phase space, are finally neglected, so that one is left with a perturbative Hamiltonian of the form

$$H = \sum_{k,M,P,N} k_{kMPN} \, n_1^k (\cos\gamma)^M \sigma^P (J^2)^N \qquad (3)$$

where the $k_{kMPN}$ are real coefficients and $n_1 = a_1^+ a_1$ denotes the stretch quantum number. The eigenvalues and eigenvectors of the Hamiltonian of Eq. (3) are easily obtained from the diagonalization of matrices of approximate size 60*60 built in the basis of the free rotor. It is reminded, that application of CPT leads to so-called "asymptotic" series, which means that the eigenvalues and eigenvectors of the Hamiltonian of Eq. (3) converge towards those of the exact Hamiltonian up to a certain order and then diverge again. The order at which the series diverges is connected to the famous small divisor problem and cannot be determined *a priori*. As can be seen in Fig. 1, there exist two MEPs in the case where $C \leq D$ : the first one is $\gamma = 0$ (the horizontal axis), while the second one, shown as a dot-dashed line in Fig. 1, can be expressed as $R_{MEP} = R_{MEP}(\gamma)$. It turns out that, for $C$=5625 cm$^{-1}$, use of the two MEP results in very similar accuracies : expansion around $\gamma = 0$ leads to an absolute average error between the 98 exact and perturbative eigenvalues smaller than 10 cm$^{-1}$ for all perturbation orders comprised between (or equal to) 4 and 13, with a minimum error of 6.75 cm$^{-1}$ at order 10, while expansion around $R_{MEP}=R_{MEP}(\gamma)$ leads to an average error smaller than 10 cm$^{-1}$ for all orders comprised between 6 and 11, with a minimum error of 6.25 cm$^{-1}$ at order 9.

In contrast, the range of values, which $\gamma$ can reach, is narrower and localized above the minimum of the PES in the case where $C>D$, so that one can use the usual procedure for semi-rigid molecules [1,14]. Accordingly, the Hamiltonian is first expanded in Taylor series around the minimum of the PES and rewritten in terms of the ladder operators associated with dimensionless normal coordinates. The series of canonical transformations $H\rightarrow\exp(S)H\exp(-S)$ is then applied in order to keep in the perturbative Hamiltonian only the terms which depend on the stretch quantum number $n_1=a_1^+a_1$ and the bend quantum number $n_2=a_{2x}^+a_{2x}+a_{2y}^+a_{2y}$, where 2*x* and 2*y* denote the two components of the degenerate bending motion. The obtained polynomes in $n_1$ and $n_2$ are called Dunham expansions (see Eq. (4) below). For *C*=18750 cm$^{-1}$, the convergence of the Dunham asymptotic series is however exceedingly slow. Examination of the differences between exact and perturbative energies shows that this slow convergence is due to a non-negligible 1:2 Fermi resonance between the two degrees of freedom. When taking this resonance into account in the choice of the successive operators *S*, the perturbative Hamiltonian is obtained in the form

$$H=\sum_{k,m}h_{km}\,n_1^k n_2^m+a_1^+\Big(\sum_{k,m}f_{km}\,n_1^k n_2^m\Big)(a_{2x}^2+a_{2y}^2)+\text{c.c.}\tag{4}$$

where the $h_{km}$ and $f_{km}$ are real coefficients. The first term in the right-hand side of Eq. (4) is the polynomial Dunham expansion and the second one the Fermi resonance. Again, the eigenvalues and eigenvectors of the Hamiltonian of Eq. (4) are very easily obtained from the diagonalization of small matrices of size less than 11*11 built in the direct product basis of the non-degenerate and doubly-degenerate harmonic oscillators. The average absolute error between the exact and perturbative energies of the 48 bound states is 12.8 cm$^{-1}$ at 10th order of CPT, but it increases again rapidly at orders larger than 13. To illustrate more clearly the agreement between exact and perturbative results, the wave functions for 3 states are shown in Fig. 2. These states (#42, #43 and #47) have respective energies 14578, 14667 and 14929 cm$^{-1}$ above the minimum of the PES, that is, they are located close to the dissociation threshold at 15000 cm$^{-1}$. Comparison of the bottom row (exact calculations) and the top one (perturbative results) shows that the wave functions are indeed very similar.

Conclusion of the first, quantum mechanical part of this study is therefore that, when the geometry of the PES is not too complex, CPT is able to provide approximations of dissociating systems with several degrees of freedom, which are simple and nevertheless precise up to the threshold. In the remainder of this Letter, we next present a few results of the classical analysis, which give a clear illustration of the simplifications brought to the investigated system by the CPT procedure.

The left column of Fig. 3 shows the $(R, p_R)$ Poincaré surfaces of section (SOS) at $\gamma$=0 and the $(\gamma, p_\gamma)$ SOS at $R$=2.5 Å for the exact Hamiltonian of Eq. (1) with $C$=5625 cm$^{-1}$ at an energy $E$=13000 cm$^{-1}$ above the minimum of the PES, that is, 2000 cm$^{-1}$ below the threshold. The $(R, p_R)$ SOS displays the usual elongated triangular form of dissociating coordinates, while the $(\gamma, p_\gamma)$ SOS looks essentially like that of the pendulum. However, at this energy, many tori have already split into resonance islands and a thin macroscopic chaotic region has appeared around the separatrix between rotation-like and vibration-like motions, the size of which increases rapidly when energy approaches the dissociation limit. The right column of Fig. 3 shows the corresponding surfaces of section at the same energy for the transformed Hamiltonian of Eq. (3), that is, the $(q_1, p_1)$ SOS at $\gamma$=0 and the $(\gamma, p_\gamma)$ SOS at $q_1$=0. The transformed Hamiltonian being separable, one should not be surprised to notice that all the subtleties (resonance islands, chaos) of the preceding SOS have disappeared. Moreover, as we already stated, the canonical transformation maps the Morse oscillator on a slightly anharmonic one. Therefore, the two SOS on the right-hand side of Fig. 3 can hardly be distinguished from those of the harmonic oscillator and the pendulum, respectively. Nonetheless, it is emphasized that the essential geometrical features of the original Hamiltonian are preserved. For example, the periods and action integrals of the periodic orbits [R], [γ] and [SN], which organize the classical phase space and also act as backbones for the quantum mechanical wave functions [5,6] (see below), agree almost perfectly for the two systems.

Fig. 4 displays the same information for the the exact Hamiltonian of Eq. (1), with $C$ however equal to 18750 $cm^{-1}$, and the perturbative Hamiltonian of Eq. (4). More precisely, the right column of Fig. 3 shows the $(q_1, p_1)$ SOS at $q_2$=0 and the $(q_2, p_2)$ SOS at $q_1$=0. In addition to the fact that $\gamma$ remains localized in the well centred around $\gamma$=0, the essential difference between Figs. 3 and 4 results from the period-doubling bifurcation, which the [R] periodic orbit (PO) undergoes around 10000 $cm^{-1}$. At this bifurcation, the [R] PO becomes unstable - we therefore label it [R*] above the bifurcation's energy - while the same PO covered twice, which is labelled [2R], remains stable and progressively acquires, with increasing energies, a pronounced ⊃ shape in the $(R,\gamma)$ plane. This bifurcation is due to the 1:2 Fermi resonance between the stretching and bending degrees of freedom and has been observed, for example, in the spectra of $CO_2$ and $CS_2$. Macroscopic chaos first appears around the separatrix encompassing [R*] and develops more rapidly than in Fig. 3, because of the higher value of the coupling coefficient $C$. It is again observed, in the right column of Fig. 4, that CPT drastically simplifies the appearance of the SOS, although it preserves the essential organization of the classical phase space around the [R*], [2R] and [γ] POs. Let us finally note that these three POs have been plotted in Fig. 2 on top of the quantum mechanical wave functions for states #42, #43 and #47. One can check, as usual, that the POs, in addition to organizing the classical phase space, also play the role of backbones for the quantum wave functions. The agreement between exact and perturbative quantum results suggests that the quantum world is much more sensitive to the properties of these POs than to the other details (resonance islands, chaos, etc...) of the classical mechanics.

In conclusion, we have shown that CPT is able to provide a fairly precise, though not exact, approximation of the Hamiltonian of a vibrating molecular system with several degrees of freedom up to the dissociation threshold. The method probably fails for too complex PES and/or too strong couplings but it is still expected to be very useful for studying the highly excited dynamics of a large variety of molecules.

## FIGURE CAPTIONS

**Figure 1** : Contour plots of the potential energy surface of Eq. (1) for coupling constants $C$=5625 cm$^{-1}$ (top plot) and $C$=18750 cm$^{-1}$ (bottom plot). Contour lines range from 1000 cm$^{-1}$ to 16000 cm$^{-1}$ above the minimum of the PES, with increments of 1000 cm$^{-1}$. Filled circles indicate the positions of the minimum of the PES (on the $\gamma$=0 axis), as well as the saddle (on the $\gamma$=$\pi$ axis) for $C$=5625 cm$^{-1}$. In this later plot, the dot-dashed line shows the $R_{MEP} = R_{MEP}(\gamma)$ Minimum Energy Path between the two equilibria. The dissociation threshold is located at 15000 cm$^{-1}$ above the bottom of the PES and the saddle at around 9140 cm$^{-1}$.

**Figure 2** : Bottom : Contour plots in the ($R$,$\gamma$) plane of the wave functions of states #42 ($E$=14578.26 cm$^{-1}$), #43 ($E$=14667.04 cm$^{-1}$) and #47 ($E$=14928.51 cm$^{-1}$) for the Hamiltonian of Eq. (1) with $C$=18750 cm$^{-1}$. The thick lines indicate the positions of the [γ], [R*] and [2R] periodic orbits at corresponding energies (the [R*] PO lies on top of the $R$ axis) and the dot-dashed lines the contours of the PES at $E$=5000, 10000 and 14900 cm$^{-1}$. Top : Same plots, in the $(q_1, q_2)$ plane, for the Hamiltonian of Eq. (4) obtained from 10th order CPT.

**Figure 3** : Left column : $(R, p_R)$ SOS at $\gamma$=0 and $(\gamma, p_\gamma)$ SOS at $R$=2.5 Å for the Hamiltonian of Eq. (1) with $C$=5625 cm$^{-1}$ at $E$=13000 cm$^{-1}$ (that is, 2000 cm$^{-1}$ below the threshold). Filled circles indicate the positions of some periodic orbits. Right column : $(q_1, p_1)$ SOS at $\gamma$=0 and $(\gamma, p_\gamma)$ SOS at $q_1$=0 for the 8th order perturbative Hamiltonian of Eq. (3) at the same energy.

**Figure 4** : Left column : $(R, p_R)$ SOS at $\gamma$=0 and $(\gamma, p_\gamma)$ SOS at $R$=2.5 Å for the Hamiltonian of Eq. (1) with $C$=18750 cm$^{-1}$ at $E$=13000 cm$^{-1}$ (that is, 2000 cm$^{-1}$ below the threshold). Filled circles indicate the positions of some periodic orbits. Right column : $(q_1, p_1)$ SOS at $q_2$=0 and $(q_2, p_2)$ SOS at $q_1$=0 for the 10th order perturbative Hamiltonian of Eq. (4) at the same energy.

FIGURE 1

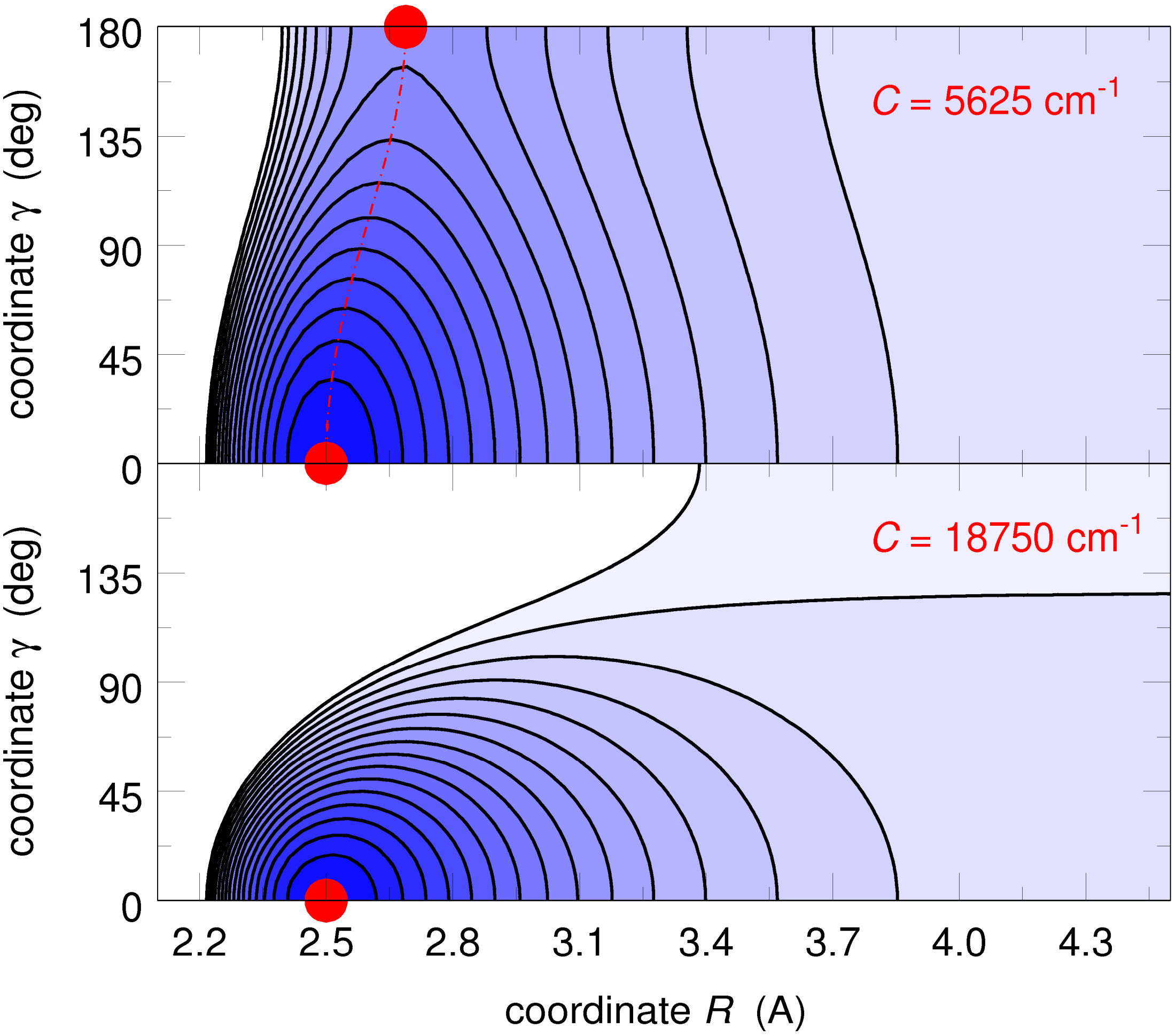

FIGURE 2

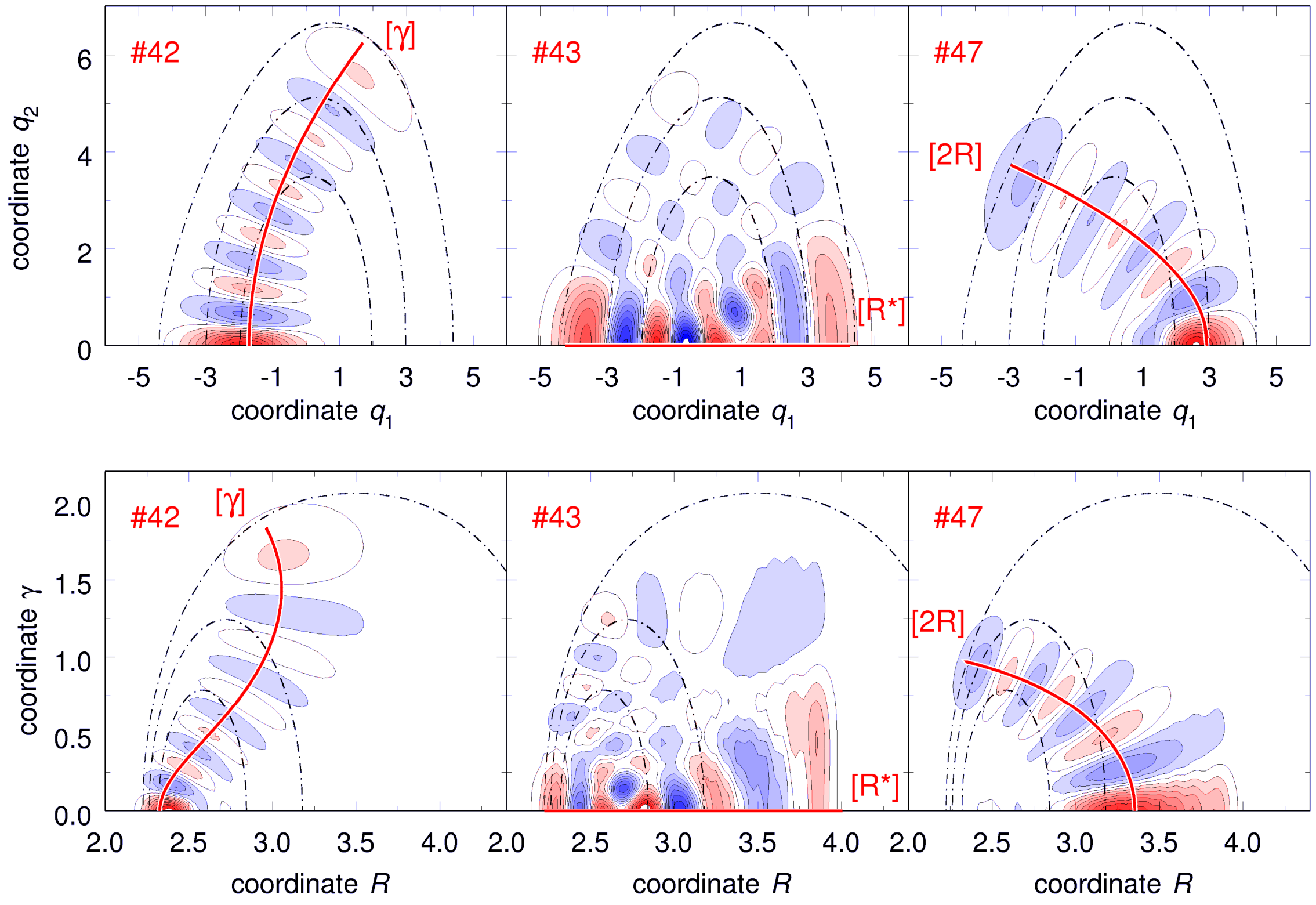

FIGURE 3

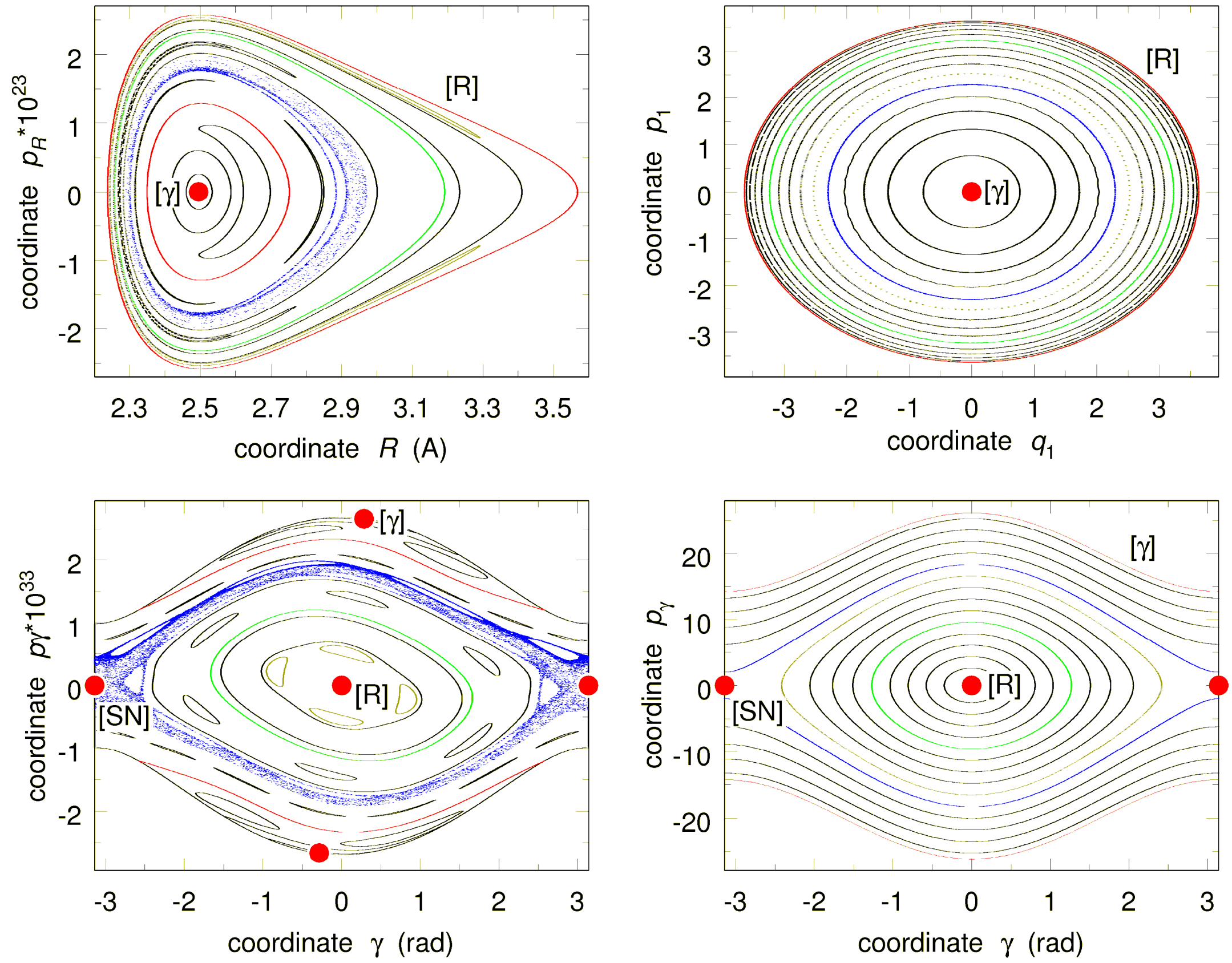

FIGURE 4

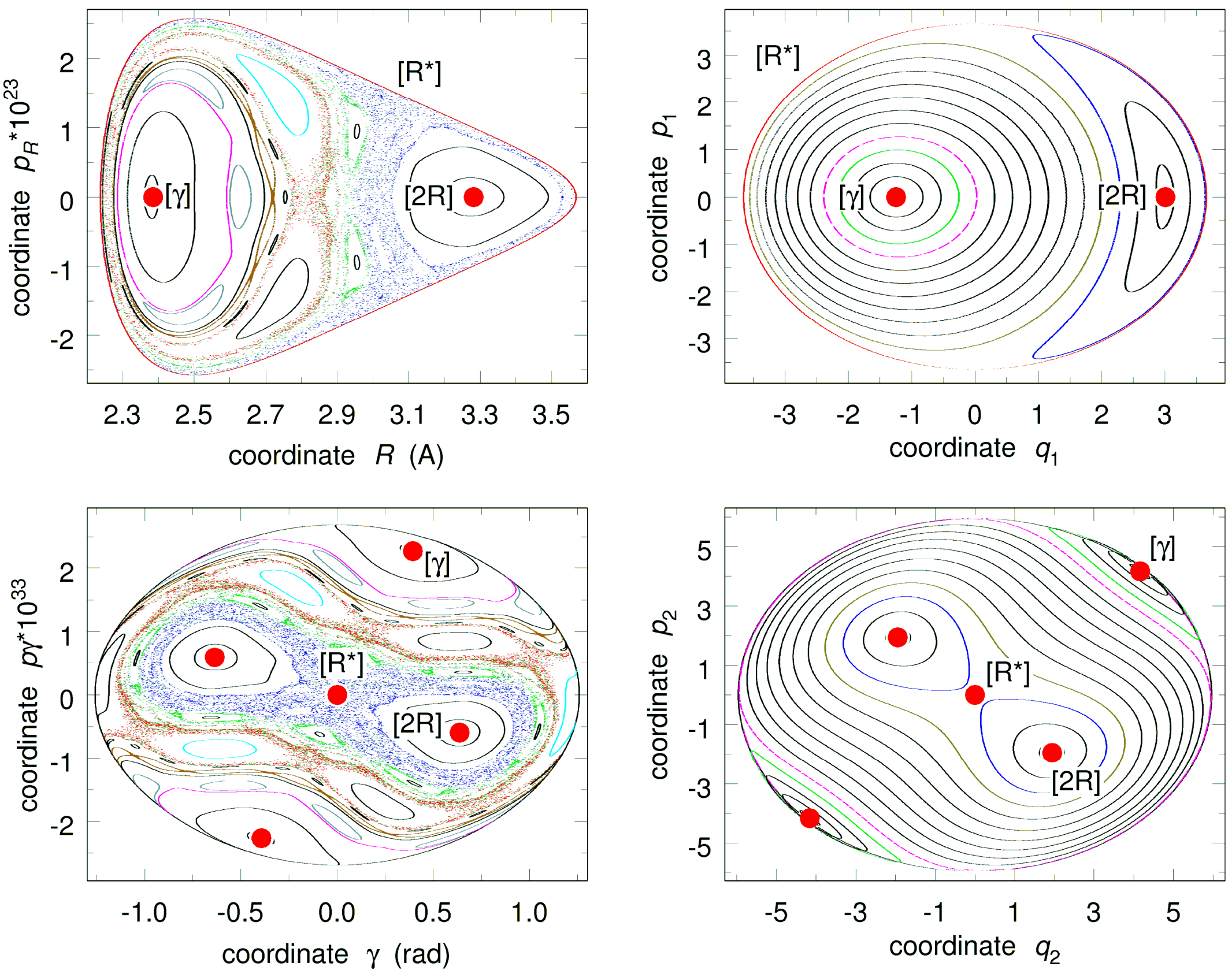